\documentclass[
 reprint,
 amsmath,amssymb,
 aps,
]{revtex4-1}

\usepackage{graphicx}
\usepackage{dcolumn}
\usepackage{bm}
\usepackage{hyperref}
\usepackage{xcolor}

\newcommand*{\bfrac}[2]{\genfrac{}{}{0pt}{}{#1}{#2}}

\begin{document}

\title{Solutions of the massive Dirac equation in the near-horizon metric of the extremal five dimensional Myers-Perry black hole with equal angular momenta}

\author{Jose Luis Bl\'azquez-Salcedo}
\email[]{jose.blazquez.salcedo@uni-oldenburg.de}
\author{Christian Knoll}
\email[]{christian.knoll@uni-oldenburg.de}
\affiliation{Institut f\"ur Physik, Universit\"at Oldenburg, D-26111 Oldenburg, Germany}

\date{\today}

\begin{abstract}
We study massive Dirac fields in the background of the near-horizon limit of the extremal Myers-Perry black hole in five dimensions. We consider the case in which both angular momenta have equal magnitude. The resulting Dirac equation can be decoupled into an angular and a radial part. The solution of the angular part results in some algebraic relations that determine completely the angular quantum numbers of the fermionic field. The radial part can be analytically solved in terms of special functions, which allow us to analyze the near-horizon radial current of the Dirac field.
\end{abstract}


\maketitle

\section{Introduction}


The study of gravity theories in higher dimensions, and in particular black hole solutions, have recently gained considerable interest. Apart from presenting interesting features, like stationary solutions with more than one plane of rotation, or non-spherical topology of the horizon \cite{Emparan:2008eg,Horowitz:2012nnc}, these solutions appear at the center of various proposals for quantum gravity, such as string theory or brane-world scenarios. Of particular interest is the 5 dimensional case, since the AdS/CFT correspondence posits a duality between solutions of 5d gravity theories and strong coupling states of 4d field theories \cite{ammon_erdmenger_2015}. 

   

An important example of a higher dimensional solution is the stationary black hole known as the Myers-Perry (MP) solution \cite{Myers:1986un}. This solution is a generalization to arbitrary dimensions of the 4d Kerr black hole, but instead of a single angular momentum, the black hole rotates in $\lfloor\frac{D-1}{2}\rfloor$ independent planes, and the horizon retains the spherical topology (in contrast with the black rings with a torus-like topology of the horizon \cite{Emparan:2001wn}). 

The Myers-Perry black hole possesses an extremal limit, similar to the Kerr metric. On the other hand specially interesting is the case when all the angular momenta have equal magnitude, because in odd dimensions the solution acquires additional symmetry (cohomogeneity-1). This particular case also possesses an extremal case \cite{Myers:1986un}.

An interesting limit of this cohomogeneity-1 extremal solution is found if one looks at the near-horizon neighborhood of the extremal case \cite{Bardeen:1999px}. In this case the isometries of the metric are enhanced even further to $AdS_2 \times S^3$, with the metric acquiring a very simple expression.  
The properties of the near-horizon geometry of the MP black hole and other more general extremal black holes have been extensively studied in the literature, and we refer the reader to \cite{Astefanesei:2006dd,Kunduri:2007vf,Kunduri:2013ana}. Although the near-horizon solution allows to calculate some properties of the global solutions, it has been also observed that the near-horizon metric does not uniquely relates to a global solution (see for example \cite{Blazquez-Salcedo:2015kja,Blazquez-Salcedo:2017kig,Blazquez-Salcedo:2017ghg}).

The interaction of black holes with test matter fields has also attracted increasing interest lately, in particular, the properties of massive fermionic fields around stationary black holes. In contrast to bosonic fields, Dirac fields lack zero modes \cite{Batic:2017qcr,Lasenby:2002mc,Daude:2012wq} or superradiant instabilities \cite{Maeda:1976tm,Iyer:1978du,Brito:2015oca}, resulting in
stationary fermionic fields around a black hole not being possible in GR, even if one allows for the presence of additional structure \cite{Finster:1998ju,Finster:1998ak,Finster:1999ry,Finster:2002vk} (although there are examples of self-gravitating and stationary configurations that make use of more than one fermionic fields to bypass this situation \cite{Herdeiro:2017fhv}).

In addition to these results, the quasinormal mode analysis of the Dirac equation in the background of several black hole spacetimes explicitly shows that these perturbations are always damped in time \cite{Jin:1998rg,Cardoso:2001hn,Jing:2003wq,Zhidenko:2003wq,Cho:2003qe,Jing:2005dt,Cho:2007zi,Arnold:2013gka,Cotaescu:2014jca,Dolan:2015eua,Cotaescu:2016aty}. However, it has been found recently that such modes can be very slowly damped if a massive fermionic field is allowed \cite{Blazquez-Salcedo:2017bld,Konoplya:2017tvu}, leading to effectively stable configurations, at least at the level of a test field (effective stability has also been discussed in the context of scalar fields \cite{Degollado:2018ypf}).

Analytical solutions to the Dirac equation can be obtained in several backgrounds \cite{Kraniotis:2018zmh,doi:10.1063/1.2912725}. In particular, the increased symmetry of the extremal black hole in the near-horizon geometry allows to study analytically several properties of the Dirac perturbation. For instance, the massive Dirac field was previously studied in the near horizon geometry of the extremal Kerr black hole in \cite{Sakalli:2004bx}.

In 5 dimensions, since the Dirac equation can be separated into an angular and a radial part for the Myers-Perry metric due to the existence of Killing-Yano tensors \cite{Oota:2007vx, Cariglia:2011qb, Frolov:2017kze, Wu:2008df}, it is expected that in the near horizon limit the separation also works.

In this paper we consider the case of a massive spin 1/2 test field in the near-horizon geometry of a 5 dimensional Myers-Perry black hole.
The paper is organized as follows: in section \ref{S1} we will write down the Dirac equation in the near horizon geometry of the five dimensional extremal Myers-Perry black hole with equal angular momenta. We will also decouple the Dirac equation into angular and radial equations. In section \ref{S2} we will solve the angular equation imposing the physically relevant boundary conditions. The solution to the radial equations is presented in section \ref{S3}. The behavior of the sign of the radial current for some simple cases near the horizon is discussed in section \ref{S4}. Lastly in section \ref{S5} we finish with some final remarks and a short discussion regarding future work.


\section{The radial and angular equations}\label{S1}

The metric of the extremal five dimensional Myers Perry black hole with equal angular momenta is \cite{Myers:1986un}
\begin{eqnarray}
\mathrm d s^2 &=& \mathrm d T^2 - \frac{\rho^2 R^2}{\Delta^2} \mathrm d \rho^2 \nonumber \\
&& - \frac{4 a^2}{R^2} \left( \mathrm d T + a \sin^2 \theta \, \mathrm d \phi_1 + a \cos^2 \theta \, \mathrm d \phi_2 \right)^2 \nonumber \\
&&  - R^2 \left( \mathrm d \theta^2 + \sin^2 \theta \, \mathrm d \phi_1^2 + \cos^2 \theta \, \mathrm d \phi_2^2 \right) \, ,
\end{eqnarray}
where we have defined the functions
\begin{eqnarray}
R^2 &=& \rho^2 + a^2 \, , \nonumber \\
\Delta &=& \rho^2 - a^2 \, .
\end{eqnarray}
Note that the metric is determined by a single parameter, $a$, which determines both the mass and the angular momentum of the black hole.

To obtain the near-horizon metric, we can make the following coordinate change
\begin{eqnarray}
\rho &=& a + \epsilon r \, , \nonumber \\
T &=& \frac{2 t}{\epsilon} \, , \nonumber \\
\phi_j &=& \psi_j - \frac{T}{2 a} \, ,
\end{eqnarray}
which introduces the scaling parameter $\epsilon$, and jumps into a frame corrotating with the horizon.

Taking the limit $\epsilon \rightarrow 0$ and only keeping the lowest order terms gives the near-horizon metric \cite{Bardeen:1999px}
\begin{eqnarray}
\mathrm d s^2 &=& \frac{2 r^2}{a^2} \mathrm d t^2 - \frac{a^2}{2 r^2} \mathrm d r^2 - 2 a^2 \mathrm d \theta^2 \nonumber \\
&& - 4 \left( \frac{r}{a} \mathrm d t - a [\sin^2 \theta \, \mathrm d \psi_1 + \cos^2 \theta \, \mathrm d \psi_2] \right)^2 \nonumber \\
&& - 2 a^2 \sin^2 \theta \, \cos^2 \theta \, (\mathrm d \psi_1 - \mathrm d \psi_2)^2 \, .
\label{NH_metric}
\end{eqnarray}
Note the near-horizon metric is a solution of the GR equations, and it depends again on a single parameter $a$ (the scaling parameter $\epsilon$ drops out of the leading term).

We are interested in solutions of the Dirac equation on this geometry
\begin{eqnarray}
\mathcal D \Psi = m \Psi \, ,
\end{eqnarray}
where $\Psi$ is the spinor, $\mathcal D$ is the Dirac operator, and $m$ is the mass of the field.
In order to introduce the Dirac field in the space-time (\ref{NH_metric}), we will use the following vielbein
\begin{eqnarray}
\boldsymbol{\omega}^0 &=& \frac{\sqrt{2} r}{a} \mathbf d t \, , \nonumber \\
\boldsymbol{\omega}^1 &=& \frac{a}{\sqrt{2} r} \mathbf d r \, , \nonumber \\
\boldsymbol{\omega}^2 &=& \sqrt{2} a \mathbf d \theta \, , \nonumber \\
\boldsymbol{\omega}^3 &=& \sqrt{2} a \sin \theta \, \cos \theta \, (\mathbf d \psi_1 - \mathbf d \psi_2) \, , \nonumber \\
\boldsymbol{\omega}^4 &=& 2 \left( a [\sin^2 \theta \, \mathbf d \psi_1 + \cos^2 \theta \, \mathbf d \psi_2 ] - \frac{r}{a} \mathbf d t \right) \, .
\end{eqnarray}
This results in the Dirac operator
\begin{eqnarray}
\mathcal D &=& \frac{\mathrm i}{\sqrt{2} a} \gamma^0 \left( \frac{a^2}{r} \partial_t + \partial_{\psi_1} + \partial_{\psi_2} \right) \nonumber \\
&&+ \frac{\sqrt{2} \mathrm i r}{a} \gamma^1 \left( \partial_r + \partial_r \ln \sqrt{r} \right)  \nonumber \\
&& + \frac{\mathrm i}{\sqrt{2} a} \gamma^2 \left( \partial_\theta + \partial_\theta \ln \sqrt{\sin \theta \, \cos \theta} \right) \nonumber \\
&&+ \frac{\mathrm i}{\sqrt{2} a} \gamma^3 (\cot \theta \, \partial_{\psi_1} - \tan \theta \, \partial_{\psi_2} ) \nonumber \\
&& + \frac{\mathrm i}{2 a} \gamma^4 (\partial_{\psi_1} + \partial_{\psi_2}) - \frac{\mathrm i}{2 a} \left( \gamma^0 \gamma^1 + \gamma^2 \gamma^3 \right) \gamma^4  \, .
\end{eqnarray}
In order to simplify the Dirac equation, it is convenient to perform the following transformation to the Dirac operator:
\begin{eqnarray}
\frac{1}{\zeta} \mathrm e^{\frac{\pi}{8} \gamma^0 \gamma^1 \gamma^2 \gamma^3} (\mathcal D - m) \zeta \mathrm e^{\frac{\pi}{8} \gamma^0 \gamma^1 \gamma^2 \gamma^3} =: \mathcal D_\ast \, ,
\end{eqnarray}
with $\zeta = 1 / \sqrt{r \sin\theta \, \cos \theta}$. 
This transformation is the near-horizon limit of the standard transformation used before in the literature \cite{Oota:2007vx, Cariglia:2011qb, Frolov:2017kze, Wu:2008df}, which allows to decouple the angular and radial parts.
Note that the $\gamma$ matrices satisfy
\begin{eqnarray}
\mathrm e^{\frac{\pi}{4} \gamma^0 \gamma^1 \gamma^2 \gamma^3} (\gamma^0 \gamma^1 + \gamma^2 \gamma^3) \gamma^4 = \sqrt{2} \gamma^2 \gamma^3 \gamma^4 \, .
\end{eqnarray}
In addition, for the Dirac algebra we will use a representation such that 
\begin{eqnarray}
\gamma^0 \gamma^1 \gamma^2 \gamma^3 \gamma^4 \equiv 1 \, .
\end{eqnarray}
This results in a much simpler expression for the modified Dirac operator $\mathcal D_\ast$,
\begin{eqnarray}
\sqrt{2} a \mathcal D_\ast &=& \mathrm i \gamma^0 \left( \frac{a^2}{r} \partial_t + \partial_{\psi_1} + \partial_{\psi_2} \right) + 2 \mathrm i r \gamma^1 \partial_r \nonumber \\
&&+ \mathrm i \gamma^0 \gamma^1 \mathcal K_\ast + \left(\frac{\mathrm i}{2} [\partial_{\psi_1} + \partial_{\psi_2}] - a m  \right)  \, .
\end{eqnarray}
We have written the Dirac operator in terms of the angular operator $\mathcal K_\ast$ 
\begin{eqnarray}
\mathcal K_\ast &=& \gamma^0 \gamma^1 \gamma^2 \partial_\theta + \gamma^0 \gamma^1 \gamma^3 (\cot \theta \, \partial_{\psi_1} - \tan \theta \, \partial_{\psi_2}) \nonumber \\
&&+ \frac{1}{2} \gamma^0 \gamma^1 \gamma^4 (\partial_{\psi_1} + \partial_{\psi_2} - 2 \mathrm i a m) - \frac{1}{2}   \, .
\end{eqnarray}
Note that both operators commute, $[\mathcal K_\ast, \mathcal D_\ast] = 0$. Hence we can write the spinor $\Psi$ like 
\begin{eqnarray}
\Psi_\ast &=& \frac{1}{\zeta} \mathrm e^{-\frac{\pi}{8} \gamma^0 \gamma^1 \gamma^2 \gamma^3} \Psi  \\
&=& \phi(r) \otimes \Theta(\theta) \, \mathrm e^{- \mathrm i \omega t + \mathrm i m_1 \psi_1 + \mathrm i m_2 \psi_2} \, . \nonumber
\label{spinor_ansatz}
\end{eqnarray}
$\phi(r)$ is the radial part of the function, and $\Theta(\theta)$ is the eigenfunction of the angular operator
\begin{eqnarray}
\mathcal K_\ast \Theta \mathrm e^{\mathrm i m_1 \psi_1 + \mathrm i m_2 \psi_2} = \kappa \Theta \mathrm e^{\mathrm i m_1 \psi_1 + \mathrm i m_2 \psi_2} \, ,
\end{eqnarray}
where the angular quantum numbers are $\kappa$, $m_1$ and $m_2$.

As a result, the angular equation can be written like
\begin{eqnarray}
\bigg\{ \hat \gamma^2 \frac{\mathrm d}{\mathrm d \theta} + \mathrm i (m_1 \cot \theta - m_2 \tan \theta) \hat \gamma^3 && \nonumber \\
 + \frac{\mathrm i (\lambda - 2 a m)}{2} \hat \gamma^4 - \frac{1}{2} & \bigg\} & \Theta = \kappa \Theta \, ,
 \label{angular_eq}
\end{eqnarray}
where we have defined 
\begin{eqnarray}
\lambda &=& m_1 + m_2 \, ,
\end{eqnarray}
and the angular $\gamma$-matrices 
\begin{eqnarray}
\hat \gamma^j &=& \gamma^0 \gamma^1 \gamma^j \; , \; j \in \{ 2, 3, 4\} \, .
\label{angular_gammas}
\end{eqnarray}

In addition, the radial part results in
\begin{eqnarray}
\left\{ 2 \mathrm i r \gamma^1 \frac{\mathrm d}{\mathrm d r} + \gamma^0 \left( \frac{\omega a^2}{r} - \lambda \right) + \mathrm i \kappa \gamma^0 \gamma^1 - \frac{\lambda + 2 a m}{2} \right\} \phi &=& 0 \, . \nonumber \\
&&\; 
 \label{radial_eq}
\end{eqnarray}

\section{Solutions of the angular equation}\label{S2}

Let us now focus on the angular part of the spinor. Because of the form of the equation (\ref{angular_eq}), it is convenient to choose the following representation for the angular $\gamma$-matrices defined in equation (\ref{angular_gammas}):
\begin{eqnarray}
\hat \gamma^2 &=& \left[ \begin{array}{cc} 0 & 1 \\ -1 & 0 \end{array} \right] \; , \; \hat \gamma^3 = \left[ \begin{array}{cc} 0 & \mathrm i \\ \mathrm i & 0 \end{array} \right] \, , \nonumber \\
\hat \gamma^4 &=& - \hat \gamma^2 \hat \gamma^3 = \left[ \begin{array}{cc} - \mathrm i & 0 \\ 0 & \mathrm i \end{array} \right] \; , \; \Theta = \left[ \begin{array}{c} \Theta_1 \\ \Theta_2 \end{array} \right] \, .
\end{eqnarray}
With this choice, equation (\ref{angular_eq}) becomes a system of coupled first order differential equations
\begin{eqnarray}
\left\{ \frac{\mathrm d}{\mathrm d \theta} + m_1 \cot \theta - m_2 \tan \theta \right\} \Theta_1  &=& - K_+ \Theta_2 \, , \nonumber \\
\left\{ \frac{\mathrm d}{\mathrm d \theta} - m_1 \cot \theta + m_2 \tan \theta \right\}  \Theta_2 &=& + K_- \Theta_1 \, ,
\label{angular_eqs}
\end{eqnarray}
with
\begin{eqnarray}
K_\pm = \frac{1 + 2 \kappa \pm (2 a m - \lambda)}{2} \, .
\end{eqnarray}

Note that the system of equations possesses the following symmetry:
\begin{eqnarray}
a \to -a \, , \ \
m_1 \to -m_1 \, , \ \
m_2 \to -m_2 \, , \nonumber \\ 
\Theta_1 \to \Theta_2 \, , \ \
\Theta_2 \to -\Theta_1 \, .
\label{angular_sym}
\end{eqnarray}

Let us study now possible solutions to the system of equations (\ref{angular_eqs}). 
In the following we will only be interested in solutions for which the spinor $\Psi$ is analytic in the angular variable $\theta \in [0, \pi / 2]$. This results into two physically different cases (modulo the symmetry (\ref{angular_sym})).

\begin{figure}
\begin{center}
\includegraphics[width=0.38\textwidth, angle =-90]{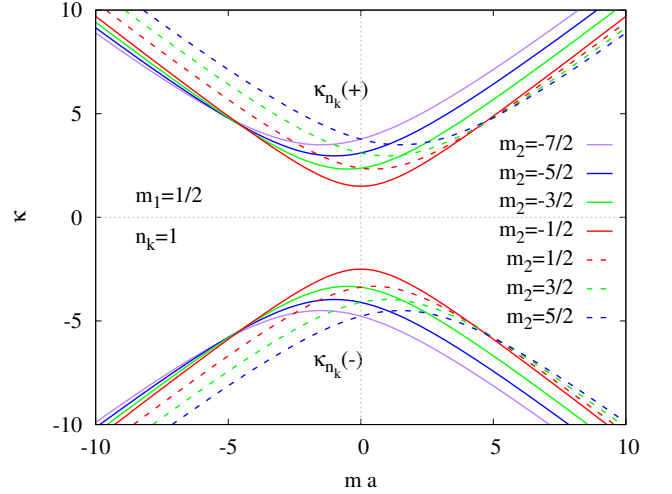}
\end{center}
\caption{
The angular eigenvalue $\kappa$ vs the product $m a$, for fixed $n_k=1$ and $m_1=1/2$. We show several values of $m_2$ and the two possible signs of $\kappa$.
}
\label{fig_kappa_case1}
\end{figure}

For the first case let us consider $K_+ \neq 0$.
We define
$p_1 := |m_1 + 1/2|$, $p_2 := |m_2 + 1/2|$
and
\begin{eqnarray}
\mathcal F_j &:=& \,_2F_1 \left( \bfrac{j + 1 - n_\kappa, j + n_\kappa + p_1 + p_2}{j + 1 + p_2} ; \cos^2 \theta \right) \, , \nonumber \\
\mathcal R_j &:=& \frac{(j+1-n_\kappa)(j+n_\kappa + p_1 + p_2)}{j + 1 + p_2} \, ,
\end{eqnarray}
where $\,_2F_1(a, b; c; z)$ is the hypergeometric function, $n_\kappa \ge 1$ is a natural number and $m_1$ and $m_2$ are half integer numbers. With these definitions the solution 
can be written like
\begin{eqnarray}
\Theta_1 &=& (\cos \theta)^{(|m_2 + 1/2| + 1/2)} \, (\sin \theta)^{(|m_1 + 1/2| + 1/2)} \, \mathcal F_0 \, , \nonumber \\
\Theta_2 &=& \left\{ \frac{2 \cos \theta \sin \theta \, \mathcal R_0 \mathcal F_1}{\mathcal F_0} - \left( m_1 + \frac{1}{2} + \left| m_1 + \frac{1}{2} \right| \right) \cot \theta \right.  \nonumber \\
&&\left. \;\; + \left( m_2 + \frac{1}{2} + \left| m_2 + \frac{1}{2} \right| \right) \tan \theta \right\} \frac{\Theta_1}{K_+} 
\label{ang_sol_case1}
\end{eqnarray}
and the angular eigenvalue is
\begin{eqnarray}
\kappa &=& - \frac{1}{2} \pm \sqrt{ \left( \frac{\lambda - 2 a m}{2} \right)^2 - \lambda^2 +\Lambda^2} \equiv \kappa_{n_k}(\pm) \, , \nonumber \\
\Lambda &=& 2 n_\kappa - 1 + |m_1 + 1/2| + |m_2 + 1/2| \,.
\label{kappa_case1}
\end{eqnarray}

Note that, apart from $m_1$, $m_2$ and $n_k$, the eigenvalue only depends on the product $m \times a$. In Figure \ref{fig_kappa_case1} we show $\kappa$ as a function of $m \times a$ for $n_k=1$, $m_1=1/2$, and several values of $m_2$. On the other hand the quantum number $n_k$ is related with the number of nodes present in the functions (\ref{ang_sol_case1}), with the number of nodes growing as this number increases. For large values of $n_k$, the angular eigenvalue behaves like $\kappa\sim \pm\sqrt{2}n_k$.

The solution is regular at $\theta=0$, where
\begin{eqnarray}
\Theta_1 \sim \theta^{|m_1+1/2|+1/2} \, , \nonumber \\
\Theta_2 \sim \theta^{|m_1-1/2|+1/2} \, , 
\end{eqnarray}
and at $\theta=\pi/2$, where 
\begin{eqnarray}
\Theta_1 \sim (\theta-\pi/2)^{|m_2+1/2|+1/2} \, , \nonumber \\
\Theta_2 \sim (\theta-\pi/2)^{|m_2-1/2|+1/2} \, . 
\end{eqnarray}

{
A second set of solutions can be obtained corresponding to the special case when $K_+ = 0$ . The solution becomes much simpler, with
\begin{eqnarray}
\Theta_1 &=& 0 \, , \nonumber \\
\Theta_2 &=& (\sin \theta)^{m_1} (\cos \theta)^{m_2} \, , 
\label{sol_case2}
\end{eqnarray}
where $m_1 > 0$ and $m_2 > 0$ are positive half integers. 
Alternatively, we can choose to set $K_- = 0$, and obtain 
\begin{eqnarray}
\Theta_1 &=& (\sin \theta)^{-m_1} (\cos \theta)^{-m_2} \, , \nonumber \\
\Theta_2 &=& 0 \, , 
\label{sol_case2b}
\end{eqnarray}
where now $m_1 < 0$ and $m_2 < 0$ are negative half integers. 
These simple cases correspond to an angular eigenvalue given by
\begin{eqnarray}
\kappa = -\frac{1}{2} \pm \frac{\lambda-2 a m}{2}  \equiv \kappa_0(\pm)
\label{kappa_case2}
\end{eqnarray}
This set of solutions can be interpreted as the $n_k=0$ limit of the previous solution (\ref{kappa_case1}), when the angular functions are in the ground state. The angular number $\kappa$ possesses a linear dependence with the product $m \times a$.
}

\section{Solutions of the radial equation}\label{S3}

Let us discuss now the radial equation (\ref{radial_eq}).
We will start assuming that $\omega \neq 0$.
In order to simplify the equation, we can make the following change of variables:
\begin{eqnarray}
z = \frac{\mathrm i \omega a^2}{r} \, .
\end{eqnarray}
With this the radial equation (\ref{radial_eq}) becomes
\begin{eqnarray}
\left\{- (\mathrm i z + \lambda) \gamma^0 - 2 \mathrm i z \gamma^1 \frac{\mathrm d}{\mathrm d z} + \mathrm i \kappa \gamma^0 \gamma^1 - \frac{\lambda + 2 am}{2} \right\} \phi &=& 0 \, . \nonumber \\
&& \;
\end{eqnarray}
Notice that the frequency $\omega$ dropped out of the differential equation with this change of variables. Let us choose the representation
\begin{eqnarray}
\gamma^0 = \left[ \begin{array}{cc} 0 & 1 \\ 1 & 0 \end{array} \right] \; , \; \gamma^1 = \left[ \begin{array}{cc} 0 & 1 \\ -1 & 0 \end{array} \right] \; , \; \phi = \left[ \begin{array}{c} \phi_1 \\ \phi_2 \end{array} \right] \, .
\label{r_rep}
\end{eqnarray}
With this choice, the radial differential equation results in the following system of coupled first order differential equations
\begin{eqnarray}
\left( 2 \mathrm i z \frac{\mathrm d}{\mathrm d z} - \mathrm i z - \lambda \right) \phi_1 &=& - \frac{2 \mathrm i \kappa - \lambda - 2 a m}{2} \phi_2 \, , \nonumber \\
\left( 2 \mathrm i z \frac{\mathrm d}{\mathrm d z} + \mathrm i z + \lambda \right) \phi_2 &=& - \frac{2 \mathrm i \kappa + \lambda + 2 a m}{2} \phi_1 \, .
\label{eqs_radial_z}
\end{eqnarray}

Let us consider analytical solutions to this system of equations 
\footnote{Formally the system (\ref{eqs_radial_z}) possesses the symmetry 
$(a,m_1,m_2,\omega,\phi_1,\phi_2)\to(-a,-m_1,-m_2,-\omega,\phi_2,-\phi_1)$. 
However this is not independent of the representation.}.
We will assume that all the coefficients of this system are different from zero, in particular $2 \mathrm i \kappa - \lambda - 2 a m \neq 0$. In this case a general solution is given by a combination of two solutions. The first solution to the system of equations (\ref{eqs_radial_z}) can be written like
\begin{eqnarray}
\sqrt{z} \phi_1^{(1)} &=& W_{\frac{\mathrm i \lambda - 1}{2}, k}(z) \, , \nonumber \\
\sqrt{z} \phi_2^{(1)} &=& \frac{4 \mathrm i}{2 \mathrm i \kappa - \lambda - 2 a m} W_{\frac{\mathrm i \lambda + 1}{2}, k}(z) \, , \nonumber \\
k^2 &=& \frac{\kappa^2 - \lambda^2}{4} + \left[ \frac{\lambda + 2 a m}{4} \right]^2 \, ,
\label{radial_1}
\end{eqnarray}
where $W_{a, b}(z)$ is the Whittaker's $W$-function. 

Similarly, the second solution can be written like
\begin{eqnarray}
\sqrt{z} \phi_1^{(2)} &=& M_{\frac{\mathrm i \lambda - 1}{2}, k}(z) \, , \nonumber \\
\sqrt{z} \phi_2^{(2)} &=& \frac{2 \lambda - 4 \mathrm i k}{2 \mathrm i \kappa - \lambda - 2 am} M_{\frac{\mathrm i \lambda + 1}{2}, k}(z) \, ,
\label{radial_2}
\end{eqnarray}
with $k$ as before and $M_{a, b}(z)$ being the Whittaker's $M$-function.

In the special case that the coefficient $2 \mathrm i \kappa - \lambda - 2 a m = 0$, these solutions simplify, becoming for the first solution
\begin{eqnarray}
\phi_1^{(1)} &=& 0 \, , \nonumber \\ 
\phi_2^{(1)} &=& z^{\frac{\mathrm i \lambda}{2}} \mathrm e^{-\frac{z}{2}}  
\end{eqnarray}
and for the second solution
\begin{eqnarray}
\phi_1^{(2)} &=& z^{-\frac{\mathrm i \lambda}{2}} \mathrm e^{\frac{z}{2}} \, , \nonumber \\
\phi_2^{(2)} &=& \frac{2 \mathrm i \kappa + \lambda + 2 a m}{4 \lambda} z^{-\frac{\mathrm i \lambda}{2}} \,_{1}{F_1} \left( \bfrac{- \mathrm i \lambda}{1 - \mathrm i \lambda} ; z \right)  \, ,
\end{eqnarray}
where $\,_1F_1$ is the confluent hypergeometric function.
{However this set of solutions are not allowed by the conditions imposed by the solutions of the angular part on the eigenvalue $\kappa$ (for instance equations (\ref{kappa_case1}) and (\ref{kappa_case2})). Hence, they are not physically relevant.}

{Finally, we will discuss the radial equation (\ref{radial_eq}) in the case that $\omega = 0$. It is convenient to make the following change of variables
\begin{eqnarray}
\xi = \ln r / a \, .
\end{eqnarray}
The radial equation (\ref{radial_eq}) then reads
\begin{eqnarray}
\left\{ 2 \mathrm i \gamma^1 \frac{\mathrm d}{\mathrm d \xi} - \lambda \gamma^0 + \mathrm i \kappa \gamma^0 \gamma^1 - \frac{\lambda + 2 a m}{2} \right\} \phi = 0 \, .
\end{eqnarray}
Even without fixing the representation, the solution can be simply written in terms of exponential functions,
\begin{eqnarray}
\phi = \exp \left\{ \left[ \kappa \gamma^0 + \mathrm i \frac{\lambda + 2 a m}{2} \gamma^1 - \mathrm i \lambda \gamma^0 \gamma^1 \right] \frac{\xi}{2} \right\} \phi_0 \, ,
\label{omega0_sol}
\end{eqnarray}
with $\phi_0$ some arbitrary constant spinor
\begin{eqnarray}
\phi_0 := \phi ( \xi = 0) = \phi( r = a) \, .
\end{eqnarray}
}

{
For comparison with the $\omega \neq 0$ case, let us choose the representation (\ref{r_rep}). Then the $\omega=0$ spinor is given by
a first solution of the form
\begin{eqnarray}
\phi_1^{(1)} &=& {\left(\frac{r}{a}\right)}^{-k} \, , \nonumber \\ 
\phi_2^{(1)} &=& \frac{2\lambda-4 \mathrm i k}{2 \mathrm i \kappa -\lambda-2 a m}{\left(\frac{r}{a}\right)}^{-k} \, ,
\end{eqnarray}
and a second solution of the form
\begin{eqnarray}
\phi_1^{(2)} &=&  {\left(\frac{r}{a}\right)}^{k} \, , \nonumber \\
\phi_2^{(2)} &=& \frac{2\lambda+4 \mathrm i k}{2 \mathrm i \kappa -\lambda-2 a m}{\left(\frac{r}{a}\right)}^{k}  \, .
\end{eqnarray}
}

\section{The fermionic current in the near-horizon region}\label{S4}

With the purpose of categorizing the previous solutions for the radial component, in  this section we will look at the behavior of the solutions as we approach the horizon. 
To make sense of the behavior at the horizon we should look at some relevant physical quantity, which in this case is the radial current, $j^r = \overline \Psi \gamma^1 \Psi$ flowing in or out of the horizon. This current has the following expression in terms of the rotated spinor (\ref{spinor_ansatz}) we have used to decouple the Dirac equation:
\begin{eqnarray}
j^r &=& \overline \Psi \gamma^1 \Psi = \Psi^\dagger \gamma^0 \gamma^1 \Psi \nonumber \\
&=& \zeta^2 \phi^\dagger \otimes \Theta^\dagger \mathrm e^{- \frac{\pi}{8} \gamma^0 \gamma^1 \gamma^2 \gamma^3} \gamma^0 \gamma^1 \mathrm e^{\frac{\pi}{8} \gamma^0 \gamma^1 \gamma^2 \gamma^3} \phi \otimes \Theta \nonumber \\
&=& \zeta^2 \left( \phi^\dagger \gamma^0 \gamma^1 \phi \right) \left| \Theta \right|^2 \, .
\end{eqnarray}

The horizon is approached when $r \rightarrow 0$. We will study the asymptotical behavior of the current corresponding to the different solutions we have presented in the previous section.

Let us begin discussing the solutions with $\omega \neq 0$.
The horizon is approached by $z \rightarrow \infty$ with the phase of $z$ being the phase of $\mathrm i \omega$ ($| \mathrm{ph} (z) | = |\mathrm{ph} (\mathrm i \omega)|$). 
The sign of the radial current is determined by $\phi^\dagger \gamma^0 \gamma^1 \phi$, which in our chosen representation (\ref{r_rep}) means that
\begin{eqnarray}
j^r \propto |\phi_2|^2 - |\phi_1|^2 \, .
\end{eqnarray}

Let us look at some simple cases for both sets of solutions. We will use the asymptotic behavior of the Whittaker functions given in the NIST handbook of mathematical functions \cite[Eq.~(13.14.20) and Eq.~(13.19.3)]{NIST:DLMF}. We will assume that $2 \mathrm i \kappa - \lambda - 2 a m \neq 0$, as imposed by the angular solutions. For the first set this means that given $| \mathrm{ph} (z) | = |\mathrm{ph} (\mathrm i \omega)| < 3 \pi / 2$
\begin{eqnarray}
\phi_1^{(1)} &\sim& z^{\frac{\mathrm i \lambda}{2} - 1} \mathrm e^{- \frac{z}{2}} \, , \nonumber \\
\phi_2^{(1)} &\sim& \frac{4 \mathrm i}{2 \mathrm i \kappa - \lambda - 2 a m} z^{\frac{\mathrm i \lambda}{2}} \mathrm e^{- \frac{z}{2}} \left[ 1 + \frac{\left|k + \frac{\mathrm i \lambda}{2} \right|^2}{z} \right] \, .
\label{radial_a1}
\end{eqnarray}
Thus in this case the functions $\phi_1$ decays faster than the function $\phi_2$ for $z \rightarrow \infty$ and thus $j^r|_{r \rightarrow 0} > 0$. 
This means the solution $\phi^{(1)}$ corresponds to emission of the fermionic field from the horizon (i.e. a white hole scenario).

Next is the second set of solutions. Given $|\mathrm{ph} (z)| = |\mathrm{ph} (\mathrm i \omega)| < \pi / 2$ and 
\begin{eqnarray}
\frac{\mathrm i \lambda \pm 1}{2} - k \neq - \frac{1}{2} , - \frac{3}{2}, \dots
\end{eqnarray}
we have
\begin{eqnarray}
\phi_1^{(2)} &\sim& \frac{\Gamma(1 + 2 k)}{\Gamma(1 + k - \frac{\mathrm i \lambda}{2} )} z^{-\frac{\mathrm i \lambda}{2}} \mathrm e^{\frac{z}{2}} \, , \nonumber \\
\phi_2^{(2)} &\sim& \frac{2 \lambda - 4 \mathrm i k}{2 \mathrm i \kappa - \lambda - 2 am} \frac{\Gamma(1 + 2 k)}{\Gamma(\frac{\mathrm i \lambda}{2} + k)} z^{-1 - \frac{\mathrm i \lambda}{2}} \mathrm e^{\frac{z}{2}} \, .
\label{radial_a2}
\end{eqnarray}
In this case the function $\phi_2$ grows slower than the function $\phi_1$ for $z \rightarrow \infty$ and thus $j^r|_{ r \rightarrow 0} < 0$.
Hence this implies that the solution $\phi^{(2)}$ corresponds to a solution being absorbed by black hole.

{Let us now suppose we have a configuration which is not damped nor exploding in time ($\Im{(\omega)} = 0$). Denote the sign of $\omega$ in this case by $\varepsilon$ and define $z := \mathrm i \varepsilon y$. Due to $r > 0$ we have $y > 0$. To obtain the asymptotical behavor of the Whittaker $M$-functions we use the asymptotic expansion to order $\mathcal O(z^{-1})$ given by \cite[Eq.~(13.19.2)]{NIST:DLMF}, meaning that the first solution is still given by equation (\ref{radial_a1}), but the second solution behaves like
\begin{eqnarray}
\phi_1^{(2)} &\sim& \frac{\Gamma(1 + 2 k)}{\Gamma( 1 + k - \frac{\mathrm i \lambda}{2})} z^{- \frac{\mathrm i \lambda}{2}} \mathrm e^{\frac{z}{2}} + \frac{\Gamma(1+2 k)}{z} \nonumber \\
&& \times \left[ \frac{\mathrm e^{\epsilon \left(1 + k - \frac{\mathrm i \lambda}{2} \right) \pi \mathrm i}}{\Gamma \left( k + \frac{\mathrm i \lambda}{2} \right)} z^\frac{\mathrm i \lambda}{2} \mathrm e^{-\frac{z}{2}}  - \frac{\left| k + \frac{\mathrm i \lambda}{2} \right|^2}{\Gamma \left( 1 + k - \frac{\mathrm i \lambda}{2} \right)} z^{- \frac{\mathrm i \lambda}{2}} \mathrm e^{\frac{z}{2}} \right] \, , \nonumber \\
\phi_2^{(2)} &\sim& \frac{2 \lambda - 4 \mathrm i k}{2 \mathrm i \kappa - \lambda - 2 am} \frac{\Gamma(1 + 2 k)}{\Gamma (1 + k + \frac{\mathrm i \lambda}{2})} z^{\frac{\mathrm i \lambda}{2}} \mathrm e^{-\frac{z}{2} + \epsilon \left( k - \frac{\mathrm i \lambda}{2} \right) \pi \mathrm i} \nonumber \\
&&+ \frac{2 \lambda - 4 \mathrm i k}{2 \mathrm i \kappa - \lambda - 2 a m} \frac{\Gamma (1 + 2 k)}{z} \nonumber \\
&& \times \left[ \frac{z^{- \frac{\mathrm i \lambda}{2}} \mathrm e^{\frac{z}{2}}}{\Gamma \left(k - \frac{\mathrm i \lambda}{2} \right)} + \frac{\left|k + \frac{\mathrm i \lambda}{2} \right|^2 \mathrm e^{\epsilon \left(k - \frac{\mathrm i \lambda}{2} \right) \pi \mathrm i}}{\Gamma \left(1 + k + \frac{\mathrm i \lambda}{2} \right)} z^{\frac{\mathrm i \lambda}{2}} \mathrm e^{- \frac{z}{2}} \right] \, . \nonumber \\
\end{eqnarray}
A zero mode with vanishing current on the horizon can be a combination of the form
\begin{eqnarray}
\phi^{(f)} = A\phi^{(1)} + B\phi^{(2)} \, ,
\end{eqnarray}
where $A \neq 0$ and $B \neq 0$ are the amplitudes of each component. Rewritten in terms of $r$, the flux has to order $\mathcal O(r^0)$ the expression
\begin{eqnarray}
j^r &\propto& \frac{1}{r}\left( |A C_2 + B C_4|^2 - |B C_3|^2 \right)  \nonumber \\
&& + 2 \Re \left\{ |A C_2 + B C_4|^2 D_2^\ast  - |B|^2 C_3^\ast D_3^{(1)} \right\} \nonumber \\
&& + 2 \Re \left\{ \left[ (D_4^{\ast} C_2 - C_3^\ast D_1) A + (D_4^{\ast} C_4 - C_3^\ast D_3^{(2)}) B \right] \right. \nonumber \\
&& \left. \;\;\;\;\;\;\;\;\;\; \times B^\ast \mathrm e^{-\mathrm i \alpha(r)} \right\} + \mathcal O (r) \, ,
\end{eqnarray}
where we have defined
\begin{eqnarray}
C_2 &=& \frac{4 \mathrm i}{2 \mathrm i \kappa - \lambda - 2 a m} (\mathrm i \omega a^2)^\frac{\mathrm i \lambda}{2} \, , \nonumber \\
C_3 &=& \frac{\Gamma(1 + 2 k)}{\Gamma(1 + k - \frac{\mathrm i \lambda}{2})} (\mathrm i \omega a^2)^{- \frac{\mathrm i \lambda}{2}} \, , \nonumber \\
C_4 &=& \frac{2 \lambda - 4 \mathrm i k}{2 \mathrm i \kappa - \lambda - 2 a m} \frac{\Gamma(1 + 2 k)}{\Gamma(1 + k + \frac{\mathrm i \lambda}{2} )}  (\mathrm i \omega a^2)^\frac{\mathrm i \lambda}{2} \mathrm e^{\mathrm i \varepsilon \pi k + \frac{\pi \varepsilon \lambda}{2}} \, . \nonumber \\
D_1 &=& (\mathrm i \omega a^2 )^{-1 + \frac{\mathrm i \lambda}{2}} \, , \nonumber \\
D_2 &=& \frac{\left| k + \frac{\mathrm i \lambda}{2} \right|^2}{\mathrm i \omega a^2} \, , \nonumber \\
D_3^{(1)} &=& - \frac{\Gamma (1 + 2 k)}{\Gamma \left(1 + k - \frac{\mathrm i \lambda}{2} \right)} \left| k + \frac{\mathrm i \lambda}{2} \right|^2 (\mathrm i \omega a^2)^{-1 - \frac{\mathrm i \lambda}{2}} \, , \nonumber \\
D_3^{(2)} &=& \frac{\Gamma(1 + 2 k)}{\Gamma \left( k  +\frac{\mathrm i \lambda}{2} \right)} \mathrm e^{\epsilon \left(1 + k - \frac{\mathrm i \lambda}{2} \right) \pi \mathrm i} (\mathrm i \omega a^2)^{-1 + \frac{\mathrm i \lambda}{2}} \, , \nonumber \\
D_4 &=& \frac{2 \lambda - 4 \mathrm i k}{2 \mathrm i \kappa - \lambda - 2 a m} \frac{\Gamma(1 + 2 k)}{\Gamma \left( k - \frac{\mathrm i \lambda}{2} \right)} (\mathrm i \omega a^2)^{-1 - \frac{\mathrm i \lambda}{2}} \, , \nonumber \\
\mathrm e^{\mathrm i \alpha(r)} &=& r^{\mathrm i \lambda} \mathrm e^z = \mathrm e^{\mathrm i \lambda \ln r + \frac{\mathrm i \omega a^2}{r}} \, .
\end{eqnarray}

Not all of the above defined quantities are independent of each other. Due to $k$ being either real or purely imaginary by definition, we have
\begin{eqnarray}
C_2 D_4^{\ast} - C_3^\ast D_1 = 0 = C_4 D_4^{\ast} - C_3^\ast D_3^{(2)} \, .
\end{eqnarray}
This means that the oscillating part in $\mathrm e^{-\mathrm i \alpha(r)}$ vanishes identically independent of the choice for $A$ and $B$. Also we have the identity
\begin{eqnarray}
D_3^{(1)} = C_3 D_2^\ast \, .
\end{eqnarray}
Using the above we can rewrite the current as
\begin{eqnarray}
j^r \propto \left( |A C_2 + B C_4|^2 - |B C_3|^2 \right) \left( r^{-1} + 2 \Re [D_2] \right) + \mathcal O (r) \, . \nonumber \\
\end{eqnarray}

An arbitrary combination of both solutions will have a divergent flux at the horizon. But it is also possible to have a constant current near the horizon, if we choose $|A C_2 + B C_4|^2 = |B C_3|^2$. Due to the above this current also automatically vanishes near the horizon. This condition can be fulfilled by appropriately choosing the amplitudes $A$ and $B$. Hence in principle, it is possible to obtain solutions that have zero modes with vanishing flux at the horizon at the level of the near-horizon solution. 

Let us now look at the current of the solution (\ref{omega0_sol}) in the special case in which we have a static perturbation with $\omega = 0$. Using this equation we can write the current like
\begin{eqnarray}
j^r = \zeta^2 (\phi^\dagger \gamma^0 \gamma^1 \phi) \left| \Theta \right|^2 = \zeta^2 (\phi_0^\dagger \gamma^0 \gamma^1 \phi_0) \left| \Theta \right|^2 \, .
\end{eqnarray}
Thus in this case, it is possible to have a vanishing current in the whole spacetime by choosing $\phi_0$ such that $\phi_0^\dagger \gamma^0 \gamma^1 \phi_0 = 0$.}

\section{Conclusions}\label{S5}

In this paper we have considered a massive Dirac field gravitating on the near-horizon region of the extremal 5d Myers-Perry black hole with equal angular momenta. 
On this near-horizon space-time we have constructed the Dirac operator that couples minimally the fermionic field to the corresponding metric. Using the standard method from the literature, based on a convenient internal rotation of the spinor, we decoupled the Dirac equation into an angular and a radial part.

First we have focused on the angular part of the solutions, which we have solved by requiring the standard regularity conditions to the spinor angular distribution. The resulting solutions possess an angular quantum number $\kappa$, which is characterized by the product $m \times a$, the half integer numbers $m_1$ and $m_2$, and the positive integer number $n_k$, related with the excitation level of the angular momentum.

Next we have considered the radial equation, which we have solved in terms of some special functions. 
Taking into account the conditions on $\kappa$ from the angular part, the solutions to the radial equation are constrained to two independent set of solutions. In the limit $\omega=0$, the radial functions of this static spinor are given in terms of simple exponential functions.

In order to study the physical properties of these radial solutions, we have analyzed the radial flux at the horizon. An asymptotical study of the analytical solutions reveals that one set of solutions is related to an ingoing flux at the horizon (absorption), while the other set is related to an outgoing flux (emission). 
We have shown explicitly that in the special case in which $\Im(\omega)=0$, a particular combination of the ingoing and outgoing solutions possesses a vanishing flux asymptotically as one approaches the horizon. We have also considered the case of static spinors ($\omega=0$), in which case the total flux in the near-horizon region can vanish.

As future work, 
since one now has a complete set of solutions for this particular near-horizon metric, the next natural step would be to quantize the field and then analyze the consequences of the quantization. 

It would also be interesting to consider a full quasi-normal mode analysis of the Dirac field in the background of the full 5d Myers-Perry black hole, without restricting the study to the near-horizon region. In this case it is necessary to study the behavior of the fields in the far-field region, and no analytical solutions are expected to be found in the bulk of the black hole. However, the quasi-normal mode analysis could reveal the presence of slowly damped modes, as observed before in higher dimensions in the Schwarzschild-Tangherlini black hole \cite{Blazquez-Salcedo:2017bld} and in 4d Kerr \cite{Konoplya:2017tvu}.

\section{Acknoledgements}\label{S6}

We would like to acknowledge support by the
DFG Research Training Group 1620 {\sl Models of Gravity}.
JLBS would like to acknowledge support from the DFG project BL 1553 and the COST Action CA16104 {\sl GWverse}. 

\bibliography{PaperV0.bbl}

\providecommand{\noopsort}[1]{}\providecommand{\singleletter}[1]{#1}%
\begin{thebibliography}{46}%
\makeatletter
\providecommand \@ifxundefined [1]{%
 \@ifx{#1\undefined}
}%
\providecommand \@ifnum [1]{%
 \ifnum #1\expandafter \@firstoftwo
 \else \expandafter \@secondoftwo
 \fi
}%
\providecommand \@ifx [1]{%
 \ifx #1\expandafter \@firstoftwo
 \else \expandafter \@secondoftwo
 \fi
}%
\providecommand \natexlab [1]{#1}%
\providecommand \enquote  [1]{``#1''}%
\providecommand \bibnamefont  [1]{#1}%
\providecommand \bibfnamefont [1]{#1}%
\providecommand \citenamefont [1]{#1}%
\providecommand \href@noop [0]{\@secondoftwo}%
\providecommand \href [0]{\begingroup \@sanitize@url \@href}%
\providecommand \@href[1]{\@@startlink{#1}\@@href}%
\providecommand \@@href[1]{\endgroup#1\@@endlink}%
\providecommand \@sanitize@url [0]{\catcode `\\12\catcode `\$12\catcode
  `\&12\catcode `\#12\catcode `\^12\catcode `\_12\catcode `\%12\relax}%
\providecommand \@@startlink[1]{}%
\providecommand \@@endlink[0]{}%
\providecommand \url  [0]{\begingroup\@sanitize@url \@url }%
\providecommand \@url [1]{\endgroup\@href {#1}{\urlprefix }}%
\providecommand \urlprefix  [0]{URL }%
\providecommand \Eprint [0]{\href }%
\providecommand \doibase [0]{http://dx.doi.org/}%
\providecommand \selectlanguage [0]{\@gobble}%
\providecommand \bibinfo  [0]{\@secondoftwo}%
\providecommand \bibfield  [0]{\@secondoftwo}%
\providecommand \translation [1]{[#1]}%
\providecommand \BibitemOpen [0]{}%
\providecommand \bibitemStop [0]{}%
\providecommand \bibitemNoStop [0]{.\EOS\space}%
\providecommand \EOS [0]{\spacefactor3000\relax}%
\providecommand \BibitemShut  [1]{\csname bibitem#1\endcsname}%
\let\auto@bib@innerbib\@empty
\bibitem [{\citenamefont {Emparan}\ and\ \citenamefont
  {Reall}(2008)}]{Emparan:2008eg}%
  \BibitemOpen
  \bibfield  {author} {\bibinfo {author} {\bibfnamefont {R.}~\bibnamefont
  {Emparan}}\ and\ \bibinfo {author} {\bibfnamefont {H.~S.}\ \bibnamefont
  {Reall}},\ }\href {\doibase 10.12942/lrr-2008-6} {\bibfield  {journal}
  {\bibinfo  {journal} {Living Rev. Rel.}\ }\textbf {\bibinfo {volume} {11}},\
  \bibinfo {pages} {6} (\bibinfo {year} {2008})},\ \Eprint
  {http://arxiv.org/abs/0801.3471} {arXiv:0801.3471 [hep-th]} \BibitemShut
  {NoStop}%
\bibitem [{\citenamefont {Horowitz}(2012)}]{Horowitz:2012nnc}%
  \BibitemOpen
  \bibinfo {editor} {\bibfnamefont {G.~T.}\ \bibnamefont {Horowitz}},\ ed.,\
  \href {http://www.cambridge.org/de/knowledge/isbn/item6633780} {\emph
  {\bibinfo {title} {{Black holes in higher dimensions}}}}\ (\bibinfo
  {publisher} {Cambridge Univ. Pr.},\ \bibinfo {address} {Cambridge, UK},\
  \bibinfo {year} {2012})\BibitemShut {NoStop}%
\bibitem [{\citenamefont {Ammon}\ and\ \citenamefont
  {Erdmenger}(2015)}]{ammon_erdmenger_2015}%
  \BibitemOpen
  \bibfield  {author} {\bibinfo {author} {\bibfnamefont {M.}~\bibnamefont
  {Ammon}}\ and\ \bibinfo {author} {\bibfnamefont {J.}~\bibnamefont
  {Erdmenger}},\ }\href {\doibase 10.1017/CBO9780511846373} {\emph {\bibinfo
  {title} {Gauge/Gravity Duality: Foundations and Applications}}}\ (\bibinfo
  {publisher} {Cambridge University Press},\ \bibinfo {year}
  {2015})\BibitemShut {NoStop}%
\bibitem [{\citenamefont {Myers}\ and\ \citenamefont
  {Perry}(1986)}]{Myers:1986un}%
  \BibitemOpen
  \bibfield  {author} {\bibinfo {author} {\bibfnamefont {R.~C.}\ \bibnamefont
  {Myers}}\ and\ \bibinfo {author} {\bibfnamefont {M.~J.}\ \bibnamefont
  {Perry}},\ }\href {\doibase 10.1016/0003-4916(86)90186-7} {\bibfield
  {journal} {\bibinfo  {journal} {Annals Phys.}\ }\textbf {\bibinfo {volume}
  {172}},\ \bibinfo {pages} {304} (\bibinfo {year} {1986})}\BibitemShut
  {NoStop}%
\bibitem [{\citenamefont {Emparan}\ and\ \citenamefont
  {Reall}(2002)}]{Emparan:2001wn}%
  \BibitemOpen
  \bibfield  {author} {\bibinfo {author} {\bibfnamefont {R.}~\bibnamefont
  {Emparan}}\ and\ \bibinfo {author} {\bibfnamefont {H.~S.}\ \bibnamefont
  {Reall}},\ }\href {\doibase 10.1103/PhysRevLett.88.101101} {\bibfield
  {journal} {\bibinfo  {journal} {Phys. Rev. Lett.}\ }\textbf {\bibinfo
  {volume} {88}},\ \bibinfo {pages} {101101} (\bibinfo {year} {2002})},\
  \Eprint {http://arxiv.org/abs/hep-th/0110260} {arXiv:hep-th/0110260 [hep-th]}
  \BibitemShut {NoStop}%
\bibitem [{\citenamefont {Bardeen}\ and\ \citenamefont
  {Horowitz}(1999)}]{Bardeen:1999px}%
  \BibitemOpen
  \bibfield  {author} {\bibinfo {author} {\bibfnamefont {J.~M.}\ \bibnamefont
  {Bardeen}}\ and\ \bibinfo {author} {\bibfnamefont {G.~T.}\ \bibnamefont
  {Horowitz}},\ }\href {\doibase 10.1103/PhysRevD.60.104030} {\bibfield
  {journal} {\bibinfo  {journal} {Phys. Rev.}\ }\textbf {\bibinfo {volume}
  {D60}},\ \bibinfo {pages} {104030} (\bibinfo {year} {1999})},\ \Eprint
  {http://arxiv.org/abs/hep-th/9905099} {arXiv:hep-th/9905099 [hep-th]}
  \BibitemShut {NoStop}%
\bibitem [{\citenamefont {Astefanesei}\ \emph {et~al.}(2006)\citenamefont
  {Astefanesei}, \citenamefont {Goldstein}, \citenamefont {Jena}, \citenamefont
  {Sen},\ and\ \citenamefont {Trivedi}}]{Astefanesei:2006dd}%
  \BibitemOpen
  \bibfield  {author} {\bibinfo {author} {\bibfnamefont {D.}~\bibnamefont
  {Astefanesei}}, \bibinfo {author} {\bibfnamefont {K.}~\bibnamefont
  {Goldstein}}, \bibinfo {author} {\bibfnamefont {R.~P.}\ \bibnamefont {Jena}},
  \bibinfo {author} {\bibfnamefont {A.}~\bibnamefont {Sen}}, \ and\ \bibinfo
  {author} {\bibfnamefont {S.~P.}\ \bibnamefont {Trivedi}},\ }\href {\doibase
  10.1088/1126-6708/2006/10/058} {\bibfield  {journal} {\bibinfo  {journal}
  {JHEP}\ }\textbf {\bibinfo {volume} {10}},\ \bibinfo {pages} {058} (\bibinfo
  {year} {2006})},\ \Eprint {http://arxiv.org/abs/hep-th/0606244}
  {arXiv:hep-th/0606244 [hep-th]} \BibitemShut {NoStop}%
\bibitem [{\citenamefont {Kunduri}\ \emph {et~al.}(2007)\citenamefont
  {Kunduri}, \citenamefont {Lucietti},\ and\ \citenamefont
  {Reall}}]{Kunduri:2007vf}%
  \BibitemOpen
  \bibfield  {author} {\bibinfo {author} {\bibfnamefont {H.~K.}\ \bibnamefont
  {Kunduri}}, \bibinfo {author} {\bibfnamefont {J.}~\bibnamefont {Lucietti}}, \
  and\ \bibinfo {author} {\bibfnamefont {H.~S.}\ \bibnamefont {Reall}},\ }\href
  {\doibase 10.1088/0264-9381/24/16/012} {\bibfield  {journal} {\bibinfo
  {journal} {Class. Quant. Grav.}\ }\textbf {\bibinfo {volume} {24}},\ \bibinfo
  {pages} {4169} (\bibinfo {year} {2007})},\ \Eprint
  {http://arxiv.org/abs/0705.4214} {arXiv:0705.4214 [hep-th]} \BibitemShut
  {NoStop}%
\bibitem [{\citenamefont {Kunduri}\ and\ \citenamefont
  {Lucietti}(2013)}]{Kunduri:2013ana}%
  \BibitemOpen
  \bibfield  {author} {\bibinfo {author} {\bibfnamefont {H.~K.}\ \bibnamefont
  {Kunduri}}\ and\ \bibinfo {author} {\bibfnamefont {J.}~\bibnamefont
  {Lucietti}},\ }\href {\doibase 10.12942/lrr-2013-8} {\bibfield  {journal}
  {\bibinfo  {journal} {Living Rev. Rel.}\ }\textbf {\bibinfo {volume} {16}},\
  \bibinfo {pages} {8} (\bibinfo {year} {2013})},\ \Eprint
  {http://arxiv.org/abs/1306.2517} {arXiv:1306.2517 [hep-th]} \BibitemShut
  {NoStop}%
\bibitem [{\citenamefont {Bl\'azquez-Salcedo}\ \emph
  {et~al.}(2015)\citenamefont {Bl\'azquez-Salcedo}, \citenamefont {Kunz},
  \citenamefont {Navarro-L\'erida},\ and\ \citenamefont
  {Radu}}]{Blazquez-Salcedo:2015kja}%
  \BibitemOpen
  \bibfield  {author} {\bibinfo {author} {\bibfnamefont {J.~L.}\ \bibnamefont
  {Bl\'azquez-Salcedo}}, \bibinfo {author} {\bibfnamefont {J.}~\bibnamefont
  {Kunz}}, \bibinfo {author} {\bibfnamefont {F.}~\bibnamefont
  {Navarro-L\'erida}}, \ and\ \bibinfo {author} {\bibfnamefont
  {E.}~\bibnamefont {Radu}},\ }\href {\doibase 10.1103/PhysRevD.92.044025}
  {\bibfield  {journal} {\bibinfo  {journal} {Phys. Rev.}\ }\textbf {\bibinfo
  {volume} {D92}},\ \bibinfo {pages} {044025} (\bibinfo {year} {2015})},\
  \Eprint {http://arxiv.org/abs/1506.07802} {arXiv:1506.07802 [gr-qc]}
  \BibitemShut {NoStop}%
\bibitem [{\citenamefont {Bl\'azquez-Salcedo}\ \emph
  {et~al.}(2018{\natexlab{a}})\citenamefont {Bl\'azquez-Salcedo}, \citenamefont
  {Kunz}, \citenamefont {Navarro-L\'erida},\ and\ \citenamefont
  {Radu}}]{Blazquez-Salcedo:2017kig}%
  \BibitemOpen
  \bibfield  {author} {\bibinfo {author} {\bibfnamefont {J.~L.}\ \bibnamefont
  {Bl\'azquez-Salcedo}}, \bibinfo {author} {\bibfnamefont {J.}~\bibnamefont
  {Kunz}}, \bibinfo {author} {\bibfnamefont {F.}~\bibnamefont
  {Navarro-L\'erida}}, \ and\ \bibinfo {author} {\bibfnamefont
  {E.}~\bibnamefont {Radu}},\ }\href {\doibase 10.1103/PhysRevD.97.081502}
  {\bibfield  {journal} {\bibinfo  {journal} {Phys. Rev.}\ }\textbf {\bibinfo
  {volume} {D97}},\ \bibinfo {pages} {081502} (\bibinfo {year}
  {2018}{\natexlab{a}})},\ \Eprint {http://arxiv.org/abs/1711.08292}
  {arXiv:1711.08292 [gr-qc]} \BibitemShut {NoStop}%
\bibitem [{\citenamefont {Bl\'azquez-Salcedo}\ \emph
  {et~al.}(2018{\natexlab{b}})\citenamefont {Bl\'azquez-Salcedo}, \citenamefont
  {Kunz}, \citenamefont {Navarro-L\'erida},\ and\ \citenamefont
  {Radu}}]{Blazquez-Salcedo:2017ghg}%
  \BibitemOpen
  \bibfield  {author} {\bibinfo {author} {\bibfnamefont {J.~L.}\ \bibnamefont
  {Bl\'azquez-Salcedo}}, \bibinfo {author} {\bibfnamefont {J.}~\bibnamefont
  {Kunz}}, \bibinfo {author} {\bibfnamefont {F.}~\bibnamefont
  {Navarro-L\'erida}}, \ and\ \bibinfo {author} {\bibfnamefont
  {E.}~\bibnamefont {Radu}},\ }\href {\doibase 10.1007/JHEP02(2018)061}
  {\bibfield  {journal} {\bibinfo  {journal} {JHEP}\ }\textbf {\bibinfo
  {volume} {02}},\ \bibinfo {pages} {061} (\bibinfo {year}
  {2018}{\natexlab{b}})},\ \Eprint {http://arxiv.org/abs/1711.10483}
  {arXiv:1711.10483 [gr-qc]} \BibitemShut {NoStop}%
\bibitem [{\citenamefont {Batic}\ \emph {et~al.}(2016)\citenamefont {Batic},
  \citenamefont {Nowakowski},\ and\ \citenamefont {Morgan}}]{Batic:2017qcr}%
  \BibitemOpen
  \bibfield  {author} {\bibinfo {author} {\bibfnamefont {D.}~\bibnamefont
  {Batic}}, \bibinfo {author} {\bibfnamefont {M.}~\bibnamefont {Nowakowski}}, \
  and\ \bibinfo {author} {\bibfnamefont {K.}~\bibnamefont {Morgan}},\ }\href
  {\doibase 10.3390/universe2040031} {\bibfield  {journal} {\bibinfo  {journal}
  {Universe}\ }\textbf {\bibinfo {volume} {2}},\ \bibinfo {pages} {31}
  (\bibinfo {year} {2016})},\ \Eprint {http://arxiv.org/abs/1701.03889}
  {arXiv:1701.03889 [gr-qc]} \BibitemShut {NoStop}%
\bibitem [{\citenamefont {Lasenby}\ \emph {et~al.}(2005)\citenamefont
  {Lasenby}, \citenamefont {Doran}, \citenamefont {Pritchard}, \citenamefont
  {Caceres},\ and\ \citenamefont {Dolan}}]{Lasenby:2002mc}%
  \BibitemOpen
  \bibfield  {author} {\bibinfo {author} {\bibfnamefont {A.}~\bibnamefont
  {Lasenby}}, \bibinfo {author} {\bibfnamefont {C.}~\bibnamefont {Doran}},
  \bibinfo {author} {\bibfnamefont {J.}~\bibnamefont {Pritchard}}, \bibinfo
  {author} {\bibfnamefont {A.}~\bibnamefont {Caceres}}, \ and\ \bibinfo
  {author} {\bibfnamefont {S.}~\bibnamefont {Dolan}},\ }\href {\doibase
  10.1103/PhysRevD.72.105014} {\bibfield  {journal} {\bibinfo  {journal} {Phys.
  Rev.}\ }\textbf {\bibinfo {volume} {D72}},\ \bibinfo {pages} {105014}
  (\bibinfo {year} {2005})},\ \Eprint {http://arxiv.org/abs/gr-qc/0209090}
  {arXiv:gr-qc/0209090 [gr-qc]} \BibitemShut {NoStop}%
\bibitem [{\citenamefont {Daude}\ and\ \citenamefont
  {Kamran}(2012)}]{Daude:2012wq}%
  \BibitemOpen
  \bibfield  {author} {\bibinfo {author} {\bibfnamefont {T.}~\bibnamefont
  {Daude}}\ and\ \bibinfo {author} {\bibfnamefont {N.}~\bibnamefont {Kamran}},\
  }\href {\doibase 10.1088/0264-9381/29/14/145007} {\bibfield  {journal}
  {\bibinfo  {journal} {Class. Quant. Grav.}\ }\textbf {\bibinfo {volume}
  {29}},\ \bibinfo {pages} {145007} (\bibinfo {year} {2012})},\ \Eprint
  {http://arxiv.org/abs/1202.1630} {arXiv:1202.1630 [math-ph]} \BibitemShut
  {NoStop}%
\bibitem [{\citenamefont {Maeda}(1976)}]{Maeda:1976tm}%
  \BibitemOpen
  \bibfield  {author} {\bibinfo {author} {\bibfnamefont {K.-i.}\ \bibnamefont
  {Maeda}},\ }\href {\doibase 10.1143/PTP.55.1677} {\bibfield  {journal}
  {\bibinfo  {journal} {Prog. Theor. Phys.}\ }\textbf {\bibinfo {volume}
  {55}},\ \bibinfo {pages} {1677} (\bibinfo {year} {1976})}\BibitemShut
  {NoStop}%
\bibitem [{\citenamefont {Iyer}\ and\ \citenamefont
  {Kumar}(1978)}]{Iyer:1978du}%
  \BibitemOpen
  \bibfield  {author} {\bibinfo {author} {\bibfnamefont {B.~R.}\ \bibnamefont
  {Iyer}}\ and\ \bibinfo {author} {\bibfnamefont {A.}~\bibnamefont {Kumar}},\
  }\href {\doibase 10.1103/PhysRevD.18.4799} {\bibfield  {journal} {\bibinfo
  {journal} {Phys. Rev.}\ }\textbf {\bibinfo {volume} {D18}},\ \bibinfo {pages}
  {4799} (\bibinfo {year} {1978})}\BibitemShut {NoStop}%
\bibitem [{\citenamefont {Brito}\ \emph {et~al.}(2015)\citenamefont {Brito},
  \citenamefont {Cardoso},\ and\ \citenamefont {Pani}}]{Brito:2015oca}%
  \BibitemOpen
  \bibfield  {author} {\bibinfo {author} {\bibfnamefont {R.}~\bibnamefont
  {Brito}}, \bibinfo {author} {\bibfnamefont {V.}~\bibnamefont {Cardoso}}, \
  and\ \bibinfo {author} {\bibfnamefont {P.}~\bibnamefont {Pani}},\ }\href
  {\doibase 10.1007/978-3-319-19000-6} {\bibfield  {journal} {\bibinfo
  {journal} {Lect. Notes Phys.}\ }\textbf {\bibinfo {volume} {906}},\ \bibinfo
  {pages} {pp.1} (\bibinfo {year} {2015})},\ \Eprint
  {http://arxiv.org/abs/1501.06570} {arXiv:1501.06570 [gr-qc]} \BibitemShut
  {NoStop}%
\bibitem [{\citenamefont {Finster}\ \emph {et~al.}(1999)\citenamefont
  {Finster}, \citenamefont {Smoller},\ and\ \citenamefont
  {Yau}}]{Finster:1998ju}%
  \BibitemOpen
  \bibfield  {author} {\bibinfo {author} {\bibfnamefont {F.}~\bibnamefont
  {Finster}}, \bibinfo {author} {\bibfnamefont {J.}~\bibnamefont {Smoller}}, \
  and\ \bibinfo {author} {\bibfnamefont {S.-T.}\ \bibnamefont {Yau}},\ }\href
  {\doibase 10.1007/s002200050675} {\bibfield  {journal} {\bibinfo  {journal}
  {Commun. Math. Phys.}\ }\textbf {\bibinfo {volume} {205}},\ \bibinfo {pages}
  {249} (\bibinfo {year} {1999})},\ \Eprint
  {http://arxiv.org/abs/gr-qc/9810048} {arXiv:gr-qc/9810048 [gr-qc]}
  \BibitemShut {NoStop}%
\bibitem [{\citenamefont {Finster}\ \emph
  {et~al.}(2000{\natexlab{a}})\citenamefont {Finster}, \citenamefont
  {Smoller},\ and\ \citenamefont {Yau}}]{Finster:1998ak}%
  \BibitemOpen
  \bibfield  {author} {\bibinfo {author} {\bibfnamefont {F.}~\bibnamefont
  {Finster}}, \bibinfo {author} {\bibfnamefont {J.}~\bibnamefont {Smoller}}, \
  and\ \bibinfo {author} {\bibfnamefont {S.-T.}\ \bibnamefont {Yau}},\ }\href
  {\doibase 10.1063/1.533234} {\bibfield  {journal} {\bibinfo  {journal} {J.
  Math. Phys.}\ }\textbf {\bibinfo {volume} {41}},\ \bibinfo {pages} {2173}
  (\bibinfo {year} {2000}{\natexlab{a}})},\ \Eprint
  {http://arxiv.org/abs/gr-qc/9805050} {arXiv:gr-qc/9805050 [gr-qc]}
  \BibitemShut {NoStop}%
\bibitem [{\citenamefont {Finster}\ \emph
  {et~al.}(2000{\natexlab{b}})\citenamefont {Finster}, \citenamefont {Kamran},
  \citenamefont {Smoller},\ and\ \citenamefont {Yau}}]{Finster:1999ry}%
  \BibitemOpen
  \bibfield  {author} {\bibinfo {author} {\bibfnamefont {F.}~\bibnamefont
  {Finster}}, \bibinfo {author} {\bibfnamefont {N.}~\bibnamefont {Kamran}},
  \bibinfo {author} {\bibfnamefont {J.}~\bibnamefont {Smoller}}, \ and\
  \bibinfo {author} {\bibfnamefont {S.-T.}\ \bibnamefont {Yau}},\ }\href
  {\doibase 10.1002/(SICI)1097-0312(200007)53:7<902::AID-CPA4>3.0.CO;2-4}
  {\bibfield  {journal} {\bibinfo  {journal} {Commun. Pure Appl. Math.}\
  }\textbf {\bibinfo {volume} {53}},\ \bibinfo {pages} {902} (\bibinfo {year}
  {2000}{\natexlab{b}})},\ \Eprint {http://arxiv.org/abs/gr-qc/9905047}
  {arXiv:gr-qc/9905047 [gr-qc]} \BibitemShut {NoStop}%
\bibitem [{\citenamefont {Finster}\ \emph {et~al.}(2002)\citenamefont
  {Finster}, \citenamefont {Smoller},\ and\ \citenamefont
  {Yau}}]{Finster:2002vk}%
  \BibitemOpen
  \bibfield  {author} {\bibinfo {author} {\bibfnamefont {F.}~\bibnamefont
  {Finster}}, \bibinfo {author} {\bibfnamefont {J.}~\bibnamefont {Smoller}}, \
  and\ \bibinfo {author} {\bibfnamefont {S.-T.}\ \bibnamefont {Yau}},\
  }\href@noop {} {\  (\bibinfo {year} {2002})},\ \Eprint
  {http://arxiv.org/abs/gr-qc/0211043} {arXiv:gr-qc/0211043 [gr-qc]}
  \BibitemShut {NoStop}%
\bibitem [{\citenamefont {Herdeiro}\ \emph {et~al.}(2017)\citenamefont
  {Herdeiro}, \citenamefont {Pombo},\ and\ \citenamefont
  {Radu}}]{Herdeiro:2017fhv}%
  \BibitemOpen
  \bibfield  {author} {\bibinfo {author} {\bibfnamefont {C.~A.~R.}\
  \bibnamefont {Herdeiro}}, \bibinfo {author} {\bibfnamefont {A.~M.}\
  \bibnamefont {Pombo}}, \ and\ \bibinfo {author} {\bibfnamefont
  {E.}~\bibnamefont {Radu}},\ }\href {\doibase 10.1016/j.physletb.2017.09.036}
  {\bibfield  {journal} {\bibinfo  {journal} {Phys. Lett.}\ }\textbf {\bibinfo
  {volume} {B773}},\ \bibinfo {pages} {654} (\bibinfo {year} {2017})},\ \Eprint
  {http://arxiv.org/abs/1708.05674} {arXiv:1708.05674 [gr-qc]} \BibitemShut
  {NoStop}%
\bibitem [{\citenamefont {Jin}(1998)}]{Jin:1998rg}%
  \BibitemOpen
  \bibfield  {author} {\bibinfo {author} {\bibfnamefont {W.~M.}\ \bibnamefont
  {Jin}},\ }\href {\doibase 10.1088/0264-9381/15/10/018} {\bibfield  {journal}
  {\bibinfo  {journal} {Class. Quant. Grav.}\ }\textbf {\bibinfo {volume}
  {15}},\ \bibinfo {pages} {3163} (\bibinfo {year} {1998})},\ \Eprint
  {http://arxiv.org/abs/gr-qc/0009009} {arXiv:gr-qc/0009009 [gr-qc]}
  \BibitemShut {NoStop}%
\bibitem [{\citenamefont {Cardoso}\ and\ \citenamefont
  {Lemos}(2001)}]{Cardoso:2001hn}%
  \BibitemOpen
  \bibfield  {author} {\bibinfo {author} {\bibfnamefont {V.}~\bibnamefont
  {Cardoso}}\ and\ \bibinfo {author} {\bibfnamefont {J.~P.~S.}\ \bibnamefont
  {Lemos}},\ }\href {\doibase 10.1103/PhysRevD.63.124015} {\bibfield  {journal}
  {\bibinfo  {journal} {Phys. Rev.}\ }\textbf {\bibinfo {volume} {D63}},\
  \bibinfo {pages} {124015} (\bibinfo {year} {2001})},\ \Eprint
  {http://arxiv.org/abs/gr-qc/0101052} {arXiv:gr-qc/0101052 [gr-qc]}
  \BibitemShut {NoStop}%
\bibitem [{\citenamefont {Jing}(2004)}]{Jing:2003wq}%
  \BibitemOpen
  \bibfield  {author} {\bibinfo {author} {\bibfnamefont {J.-l.}\ \bibnamefont
  {Jing}},\ }\href {\doibase 10.1103/PhysRevD.69.084009} {\bibfield  {journal}
  {\bibinfo  {journal} {Phys. Rev.}\ }\textbf {\bibinfo {volume} {D69}},\
  \bibinfo {pages} {084009} (\bibinfo {year} {2004})},\ \Eprint
  {http://arxiv.org/abs/gr-qc/0312079} {arXiv:gr-qc/0312079 [gr-qc]}
  \BibitemShut {NoStop}%
\bibitem [{\citenamefont {Zhidenko}(2004)}]{Zhidenko:2003wq}%
  \BibitemOpen
  \bibfield  {author} {\bibinfo {author} {\bibfnamefont {A.}~\bibnamefont
  {Zhidenko}},\ }\href {\doibase 10.1088/0264-9381/21/1/019} {\bibfield
  {journal} {\bibinfo  {journal} {Class. Quant. Grav.}\ }\textbf {\bibinfo
  {volume} {21}},\ \bibinfo {pages} {273} (\bibinfo {year} {2004})},\ \Eprint
  {http://arxiv.org/abs/gr-qc/0307012} {arXiv:gr-qc/0307012 [gr-qc]}
  \BibitemShut {NoStop}%
\bibitem [{\citenamefont {Cho}(2003)}]{Cho:2003qe}%
  \BibitemOpen
  \bibfield  {author} {\bibinfo {author} {\bibfnamefont {H.~T.}\ \bibnamefont
  {Cho}},\ }\href {\doibase 10.1103/PhysRevD.68.024003} {\bibfield  {journal}
  {\bibinfo  {journal} {Phys. Rev.}\ }\textbf {\bibinfo {volume} {D68}},\
  \bibinfo {pages} {024003} (\bibinfo {year} {2003})},\ \Eprint
  {http://arxiv.org/abs/gr-qc/0303078} {arXiv:gr-qc/0303078 [gr-qc]}
  \BibitemShut {NoStop}%
\bibitem [{\citenamefont {Jing}(2005)}]{Jing:2005dt}%
  \BibitemOpen
  \bibfield  {author} {\bibinfo {author} {\bibfnamefont {J.-l.}\ \bibnamefont
  {Jing}},\ }\href {\doibase 10.1103/PhysRevD.71.124006} {\bibfield  {journal}
  {\bibinfo  {journal} {Phys. Rev.}\ }\textbf {\bibinfo {volume} {D71}},\
  \bibinfo {pages} {124006} (\bibinfo {year} {2005})},\ \Eprint
  {http://arxiv.org/abs/gr-qc/0502023} {arXiv:gr-qc/0502023 [gr-qc]}
  \BibitemShut {NoStop}%
\bibitem [{\citenamefont {Cho}\ \emph {et~al.}(2007)\citenamefont {Cho},
  \citenamefont {Cornell}, \citenamefont {Doukas},\ and\ \citenamefont
  {Naylor}}]{Cho:2007zi}%
  \BibitemOpen
  \bibfield  {author} {\bibinfo {author} {\bibfnamefont {H.~T.}\ \bibnamefont
  {Cho}}, \bibinfo {author} {\bibfnamefont {A.~S.}\ \bibnamefont {Cornell}},
  \bibinfo {author} {\bibfnamefont {J.}~\bibnamefont {Doukas}}, \ and\ \bibinfo
  {author} {\bibfnamefont {W.}~\bibnamefont {Naylor}},\ }\href {\doibase
  10.1103/PhysRevD.75.104005} {\bibfield  {journal} {\bibinfo  {journal} {Phys.
  Rev.}\ }\textbf {\bibinfo {volume} {D75}},\ \bibinfo {pages} {104005}
  (\bibinfo {year} {2007})},\ \Eprint {http://arxiv.org/abs/hep-th/0701193}
  {arXiv:hep-th/0701193 [hep-th]} \BibitemShut {NoStop}%
\bibitem [{\citenamefont {Arnold}\ and\ \citenamefont
  {Szepietowski}(2013)}]{Arnold:2013gka}%
  \BibitemOpen
  \bibfield  {author} {\bibinfo {author} {\bibfnamefont {P.}~\bibnamefont
  {Arnold}}\ and\ \bibinfo {author} {\bibfnamefont {P.}~\bibnamefont
  {Szepietowski}},\ }\href {\doibase 10.1103/PhysRevD.88.086002} {\bibfield
  {journal} {\bibinfo  {journal} {Phys. Rev.}\ }\textbf {\bibinfo {volume}
  {D88}},\ \bibinfo {pages} {086002} (\bibinfo {year} {2013})},\ \Eprint
  {http://arxiv.org/abs/1308.0341} {arXiv:1308.0341 [hep-th]} \BibitemShut
  {NoStop}%
\bibitem [{\citenamefont {Cotaescu}\ \emph
  {et~al.}(2016{\natexlab{a}})\citenamefont {Cotaescu}, \citenamefont
  {Crucean},\ and\ \citenamefont {Sporea}}]{Cotaescu:2014jca}%
  \BibitemOpen
  \bibfield  {author} {\bibinfo {author} {\bibfnamefont {I.~I.}\ \bibnamefont
  {Cotaescu}}, \bibinfo {author} {\bibfnamefont {C.}~\bibnamefont {Crucean}}, \
  and\ \bibinfo {author} {\bibfnamefont {C.~A.}\ \bibnamefont {Sporea}},\
  }\href {\doibase 10.1140/epjc/s10052-016-3936-9} {\bibfield  {journal}
  {\bibinfo  {journal} {Eur. Phys. J.}\ }\textbf {\bibinfo {volume} {C76}},\
  \bibinfo {pages} {102} (\bibinfo {year} {2016}{\natexlab{a}})},\ \Eprint
  {http://arxiv.org/abs/1409.7201} {arXiv:1409.7201 [gr-qc]} \BibitemShut
  {NoStop}%
\bibitem [{\citenamefont {Dolan}\ and\ \citenamefont
  {Dempsey}(2015)}]{Dolan:2015eua}%
  \BibitemOpen
  \bibfield  {author} {\bibinfo {author} {\bibfnamefont {S.~R.}\ \bibnamefont
  {Dolan}}\ and\ \bibinfo {author} {\bibfnamefont {D.}~\bibnamefont
  {Dempsey}},\ }\href {\doibase 10.1088/0264-9381/32/18/184001} {\bibfield
  {journal} {\bibinfo  {journal} {Class. Quant. Grav.}\ }\textbf {\bibinfo
  {volume} {32}},\ \bibinfo {pages} {184001} (\bibinfo {year} {2015})},\
  \Eprint {http://arxiv.org/abs/1504.03190} {arXiv:1504.03190 [gr-qc]}
  \BibitemShut {NoStop}%
\bibitem [{\citenamefont {Cotaescu}\ \emph
  {et~al.}(2016{\natexlab{b}})\citenamefont {Cotaescu}, \citenamefont
  {Crucean},\ and\ \citenamefont {Sporea}}]{Cotaescu:2016aty}%
  \BibitemOpen
  \bibfield  {author} {\bibinfo {author} {\bibfnamefont {I.~I.}\ \bibnamefont
  {Cotaescu}}, \bibinfo {author} {\bibfnamefont {C.}~\bibnamefont {Crucean}}, \
  and\ \bibinfo {author} {\bibfnamefont {C.}~\bibnamefont {Sporea}},\ }\href
  {\doibase 10.1140/epjc/s10052-016-4260-0} {\bibfield  {journal} {\bibinfo
  {journal} {Eur. Phys. J.}\ }\textbf {\bibinfo {volume} {C76}},\ \bibinfo
  {pages} {413} (\bibinfo {year} {2016}{\natexlab{b}})},\ \Eprint
  {http://arxiv.org/abs/1601.03673} {arXiv:1601.03673 [gr-qc]} \BibitemShut
  {NoStop}%
\bibitem [{\citenamefont {Bl\'azquez-Salcedo}\ and\ \citenamefont
  {Knoll}(2018)}]{Blazquez-Salcedo:2017bld}%
  \BibitemOpen
  \bibfield  {author} {\bibinfo {author} {\bibfnamefont {J.~L.}\ \bibnamefont
  {Bl\'azquez-Salcedo}}\ and\ \bibinfo {author} {\bibfnamefont
  {C.}~\bibnamefont {Knoll}},\ }\href {\doibase 10.1103/PhysRevD.97.044020}
  {\bibfield  {journal} {\bibinfo  {journal} {Phys. Rev.}\ }\textbf {\bibinfo
  {volume} {D97}},\ \bibinfo {pages} {044020} (\bibinfo {year} {2018})},\
  \Eprint {http://arxiv.org/abs/1709.07864} {arXiv:1709.07864 [gr-qc]}
  \BibitemShut {NoStop}%
\bibitem [{\citenamefont {Konoplya}\ and\ \citenamefont
  {Zhidenko}(2018)}]{Konoplya:2017tvu}%
  \BibitemOpen
  \bibfield  {author} {\bibinfo {author} {\bibfnamefont {R.~A.}\ \bibnamefont
  {Konoplya}}\ and\ \bibinfo {author} {\bibfnamefont {A.}~\bibnamefont
  {Zhidenko}},\ }\href {\doibase 10.1103/PhysRevD.97.084034} {\bibfield
  {journal} {\bibinfo  {journal} {Phys. Rev.}\ }\textbf {\bibinfo {volume}
  {D97}},\ \bibinfo {pages} {084034} (\bibinfo {year} {2018})},\ \Eprint
  {http://arxiv.org/abs/1712.06667} {arXiv:1712.06667 [gr-qc]} \BibitemShut
  {NoStop}%
\bibitem [{\citenamefont {Degollado}\ \emph {et~al.}(2018)\citenamefont
  {Degollado}, \citenamefont {Herdeiro},\ and\ \citenamefont
  {Radu}}]{Degollado:2018ypf}%
  \BibitemOpen
  \bibfield  {author} {\bibinfo {author} {\bibfnamefont {J.~C.}\ \bibnamefont
  {Degollado}}, \bibinfo {author} {\bibfnamefont {C.~A.~R.}\ \bibnamefont
  {Herdeiro}}, \ and\ \bibinfo {author} {\bibfnamefont {E.}~\bibnamefont
  {Radu}},\ }\href {\doibase 10.1016/j.physletb.2018.04.052} {\bibfield
  {journal} {\bibinfo  {journal} {Phys. Lett.}\ }\textbf {\bibinfo {volume}
  {B781}},\ \bibinfo {pages} {651} (\bibinfo {year} {2018})},\ \Eprint
  {http://arxiv.org/abs/1802.07266} {arXiv:1802.07266 [gr-qc]} \BibitemShut
  {NoStop}%
\bibitem [{\citenamefont {Kraniotis}(2018)}]{Kraniotis:2018zmh}%
  \BibitemOpen
  \bibfield  {author} {\bibinfo {author} {\bibfnamefont {G.~V.}\ \bibnamefont
  {Kraniotis}},\ }\href@noop {} {\  (\bibinfo {year} {2018})},\ \Eprint
  {http://arxiv.org/abs/1801.03157} {arXiv:1801.03157 [gr-qc]} \BibitemShut
  {NoStop}%
\bibitem [{\citenamefont {Al-Badawi}\ and\ \citenamefont
  {Sakalli}(2008)}]{doi:10.1063/1.2912725}%
  \BibitemOpen
  \bibfield  {author} {\bibinfo {author} {\bibfnamefont {A.}~\bibnamefont
  {Al-Badawi}}\ and\ \bibinfo {author} {\bibfnamefont {I.}~\bibnamefont
  {Sakalli}},\ }\href {\doibase 10.1063/1.2912725} {\bibfield  {journal}
  {\bibinfo  {journal} {Journal of Mathematical Physics}\ }\textbf {\bibinfo
  {volume} {49}},\ \bibinfo {pages} {052501} (\bibinfo {year} {2008})},\
  \Eprint {http://arxiv.org/abs/https://doi.org/10.1063/1.2912725}
  {https://doi.org/10.1063/1.2912725} \BibitemShut {NoStop}%
\bibitem [{\citenamefont {Sakalli}\ and\ \citenamefont
  {Halilsoy}(2004)}]{Sakalli:2004bx}%
  \BibitemOpen
  \bibfield  {author} {\bibinfo {author} {\bibfnamefont {I.}~\bibnamefont
  {Sakalli}}\ and\ \bibinfo {author} {\bibfnamefont {M.}~\bibnamefont
  {Halilsoy}},\ }\href {\doibase 10.1103/PhysRevD.69.124012} {\bibfield
  {journal} {\bibinfo  {journal} {Phys. Rev.}\ }\textbf {\bibinfo {volume}
  {D69}},\ \bibinfo {pages} {124012} (\bibinfo {year} {2004})},\ \Eprint
  {http://arxiv.org/abs/gr-qc/0403061} {arXiv:gr-qc/0403061 [gr-qc]}
  \BibitemShut {NoStop}%
\bibitem [{\citenamefont {Oota}\ and\ \citenamefont
  {Yasui}(2008)}]{Oota:2007vx}%
  \BibitemOpen
  \bibfield  {author} {\bibinfo {author} {\bibfnamefont {T.}~\bibnamefont
  {Oota}}\ and\ \bibinfo {author} {\bibfnamefont {Y.}~\bibnamefont {Yasui}},\
  }\href {\doibase 10.1016/j.physletb.2007.11.057} {\bibfield  {journal}
  {\bibinfo  {journal} {Phys. Lett.}\ }\textbf {\bibinfo {volume} {B659}},\
  \bibinfo {pages} {688} (\bibinfo {year} {2008})},\ \Eprint
  {http://arxiv.org/abs/0711.0078} {arXiv:0711.0078 [hep-th]} \BibitemShut
  {NoStop}%
\bibitem [{\citenamefont {Cariglia}\ \emph {et~al.}(2011)\citenamefont
  {Cariglia}, \citenamefont {Krtous},\ and\ \citenamefont
  {Kubiznak}}]{Cariglia:2011qb}%
  \BibitemOpen
  \bibfield  {author} {\bibinfo {author} {\bibfnamefont {M.}~\bibnamefont
  {Cariglia}}, \bibinfo {author} {\bibfnamefont {P.}~\bibnamefont {Krtous}}, \
  and\ \bibinfo {author} {\bibfnamefont {D.}~\bibnamefont {Kubiznak}},\ }\href
  {\doibase 10.1103/PhysRevD.84.024008} {\bibfield  {journal} {\bibinfo
  {journal} {Phys. Rev.}\ }\textbf {\bibinfo {volume} {D84}},\ \bibinfo {pages}
  {024008} (\bibinfo {year} {2011})},\ \Eprint {http://arxiv.org/abs/1104.4123}
  {arXiv:1104.4123 [hep-th]} \BibitemShut {NoStop}%
\bibitem [{\citenamefont {Frolov}\ \emph {et~al.}(2017)\citenamefont {Frolov},
  \citenamefont {Krtous},\ and\ \citenamefont {Kubiznak}}]{Frolov:2017kze}%
  \BibitemOpen
  \bibfield  {author} {\bibinfo {author} {\bibfnamefont {V.}~\bibnamefont
  {Frolov}}, \bibinfo {author} {\bibfnamefont {P.}~\bibnamefont {Krtous}}, \
  and\ \bibinfo {author} {\bibfnamefont {D.}~\bibnamefont {Kubiznak}},\ }\href
  {\doibase 10.1007/s41114-017-0009-9} {\bibfield  {journal} {\bibinfo
  {journal} {Living Rev. Rel.}\ }\textbf {\bibinfo {volume} {20}},\ \bibinfo
  {pages} {6} (\bibinfo {year} {2017})},\ \Eprint
  {http://arxiv.org/abs/1705.05482} {arXiv:1705.05482 [gr-qc]} \BibitemShut
  {NoStop}%
\bibitem [{\citenamefont {Wu}(2008)}]{Wu:2008df}%
  \BibitemOpen
  \bibfield  {author} {\bibinfo {author} {\bibfnamefont {S.-Q.}\ \bibnamefont
  {Wu}},\ }\href {\doibase 10.1103/PhysRevD.78.064052} {\bibfield  {journal}
  {\bibinfo  {journal} {Phys. Rev.}\ }\textbf {\bibinfo {volume} {D78}},\
  \bibinfo {pages} {064052} (\bibinfo {year} {2008})},\ \Eprint
  {http://arxiv.org/abs/0807.2114} {arXiv:0807.2114 [hep-th]} \BibitemShut
  {NoStop}%
\bibitem [{Note1()}]{Note1}%
  \BibitemOpen
  \bibinfo {note} {Formally the system (\ref {eqs_radial_z}) possesses the
  symmetry $(a,m_1,m_2,\omega ,\phi _1,\phi _2)\to (-a,-m_1,-m_2,-\omega ,\phi
  _2,-\phi _1)$. However this is not independent of the
  representation.}\BibitemShut {Stop}%
\bibitem [{{\relax DLMF}()}]{NIST:DLMF}%
  \BibitemOpen
  {\relax DLMF},\ \href {http://dlmf.nist.gov/} {\enquote {\bibinfo {title}
  {{\it NIST Digital Library of Mathematical Functions}},}\ }\bibinfo
  {howpublished} {http://dlmf.nist.gov/, Release 1.0.18 of 2018-03-27},\
  \bibinfo {note} {f.~W.~J. Olver, A.~B. {Olde Daalhuis}, D.~W. Lozier, B.~I.
  Schneider, R.~F. Boisvert, C.~W. Clark, B.~R. Miller and B.~V. Saunders,
  eds.}\BibitemShut {Stop}%
\end{thebibliography}%

\end{document}